\documentstyle{article}

\newlength{\extraspace}
\newlength{\extraspaces}
\newcommand{\be}{\begin{equation}
\addtolength{\abovedisplayskip}{\extraspaces}
\addtolength{\belowdisplayskip}{\extraspaces}
\addtolength{\abovedisplayshortskip}{\extraspace}
\addtolength{\belowdisplayshortskip}{\extraspace}}
\newcommand{\ee}{\end{equation}}

\newcommand{\ba}{\begin{eqnarray}
\addtolength{\abovedisplayskip}{\extraspaces}
\addtolength{\belowdisplayskip}{\extraspaces}
\addtolength{\abovedisplayshortskip}{\extraspace}
\addtolength{\belowdisplayshortskip}{\extraspace}}
\newcommand{\ea}{\end{eqnarray}}

\newcommand{\newsection}[1]{
\vspace{7mm}
\pagebreak[3]
\addtocounter{section}{1}
\setcounter{equation}{0}
\setcounter{subsection}{0}
\large {\bf \thesection. #1}
\nopagebreak
\medskip
\nopagebreak
\hspace{3mm}}

\newcommand{\nonu}{\nonumber \\[.5mm]} 

\begin{document}       

\begin{large}
\centerline{\bf Cosmological Solution to Einstein-Vlasov System}
\end{large}

\centerline{Tigran Aivazian \footnote{Please use email 
tigran@zetnet.co.uk for correspondence}}

\centerline{\bf Abstract}

Einstein-Vlasov system is solved for a homogeneous isotropic spacetime 
with positive curvature ($R \times S^3$ topology). For the Universe 
consisting of massless particles the equation for $R(t)$ is solved 
analytically.

\newsection{General Equations}

We start with well-known system of self-consistent field equations of 
Einstein-Vlasov which, in geometrical form can be written as:
\be
Ein(g)=\int f p \otimes p d\mu
\ee
\be
Xf=0
\ee
where $X$ is a vector field which generates phase flow. We use geometrical
system of units where $c=1,G=1/8\pi$. In coordinate form the system 
(1.1) can be rewritten as:
\be
R_{\mu \nu}-1/2Rg_{\mu \nu}=
{2 \over \sqrt[]{-g}} \int f(x,p)p_\mu p_\nu \delta (p^2+m^2) d^4p
\ee
\be
g^{\mu \nu} p_\nu {\partial{f}\over{}\partial{x^\mu }} =
{1\over{2}}{\partial{g^{\nu \lambda}}\over{}\partial{x^\mu }}
p_\nu{}p_\lambda{}{\partial{f}\over{}{\partial{p_\mu }}}
\ee
where $f(x,p)$ is a distribution function of matter consisting of particles
with mass $m\geq{0}$. It is assumed to be a non-negative function with compact
support on a mass shell in cotangent bundle $T^*M$:
\ba
S_m(M)=\{(x,p)\in T^*M \mid g(p,p)+m^2=0\},\nonu
f:S_m(M) \rightarrow R_+,\nonu
S_m(M)=S_m^{+}(M)\cup S_m^{-}(M),\nonu
F_\pm = f\mid{_S{_m^{\pm}}}
\ea
Note, that our expression for 
$T_{\mu \nu}(x)$ differs from the one used in [1] only slightly, which 
allows to avoid Christoffel symbols in the equations for characteristics.
This corresponds to working in $(x^\mu , p_\nu )$ coordinates as opposed to
$(x^\mu , p^\nu )$.

It is convenient to use ADM 3+1 [2] representation to solve Cauchy problem.
In Appendix A we provide the 3+1 representation of Einstein-Vlasov system 
for a generic metric. In synchronous frame of reference, metric tensor
has the following form:
\be
g_{\mu \nu}=\pmatrix{
1 & 0 \cr
0 & \gamma{_{i k}} \cr
}
\ee
The dynamical equations are as follows:
\be
\partial{_t}\gamma{_{i k}}=-2K_{i k}
\ee
\be
\partial{_t}K_{i k}=R_{i k}-2K_{i l}K^l{}_k{}+KK_{i k}+
{1\over{2}}\gamma{_{i k}}(\gamma{^{l m}}T_{l m}-T_{0 0})-T_{i k}
\ee
Constraint equations are:
\be
R+K^2-K^{i k}K_{i k}=2 T_{0 0}
\ee
\be
D_j{}(K^j{}_i{}-\delta{^j{}_i{}}K)=-T_{0 i}
\ee
where $R= \gamma{^{i k}}K_{i k}$, $D_i{}$ is a covariant derivative in
$\gamma{_{i k}}$ metric. It is not difficult to see that kinetic equation 
for $f(x,p)$ will split into a couple of equations for $F_+(x,p,t)$ and 
$F_-(x,p,t)$ which are constructed by reducing $f(x,p)$ on "upper" and 
"lower" halves of mass shell. The reason for being able to reduce an equation
on $S_m(M)$ into a pair of equations for each connected half of it is due to
the fact of $S_m(M)$ is dynamically invariant with respect to the phase
flow generated by the vector field X. Here, we changed the meaning 
of what is implied by $(x,p)$ as it is usually trivial from context to 
guess to which space it belongs to. Explicit form of kinetic equations
becomes:
\be
\partial{_t}F_\pm=-{1\over{}p_0^\pm}
\biggl(\gamma{^{i k}}p_k{}{\partial{}\over{}\partial{x^i}}-
{1\over{2}}{\partial{}\gamma{^{k l}}\over{}\partial{x^i}}p_kp_l
{\partial{}\over{}\partial{p_i}}\biggr)F_\pm
\ee
or, in terms of 3-Hamiltonian:
\be
p_0^\pm{}\partial{_t}F_\pm{}=\{H,F_\pm{}\}
\ee
\be
H(x,p,t)={1\over{2}}\gamma{^{i k}}(x,t) p_i p_k
\ee
Presence of $p_0$ in these equations points to the fact that equations of
characteristics for reduced kinetic euqations do not necessarily constitute
a Hamiltonian system. The metric of the spacetime can be written in 
Robertson-Walker form:
\be
ds^2=-dt^2+R^2(t)[d\chi{^2}+\sin{^2\chi}(d\theta{^2}+\sin{^2\chi}d\theta{^2})]
\ee
Then it is not hard to calculate the components of energy-momentum 
tensor $T_{\mu \nu}(x)$:
\ba
k={1 \over R^3 \sin{^2\chi}\sin{\theta}},\nonu
T_{0 0}=k \int{}{} (F_+ + F_-) \sqrt[]{m^2 + {\lambda{}(p,p)\over{R^2}}}d^3p,
\nonu
T_{i k}=k \int{}{} {(F_+ + F_-) p_i p_k d^3p \over 
\sqrt[]{m^2 + {\lambda{}(p,p)\over{R^2}}}},\nonu
T_{0 i}=k \int (F_+ - F_-) p_i d^3p
\ea
where $\lambda{}(p,p)=p_1^2+p_2^2/\sin{^2\chi}+
p_3^2/(\sin{^2\chi}\sin{^2\theta})$. By the following transformation 
$\lambda$ is brought to canonical form:
\ba
k_1=p_1,\nonu
k_2=p_2/\sin{\chi},\nonu
k_3=p_3/(\sin{\chi}\sin{\theta})
\ea
In k-coordinates it is not hard to see the set of solutions for kinetic
equations:
\be
F_\pm =F_\pm (k)
\ee
where $k=k_1^2+k_2^2+k_3^2$. The first constraint equation takes the
following form:
\be
\dot R^2=
-1+{1\over{3R}} \int{}{} F(k) \sqrt[]{m^2 + k^2/R^2} d^3k
\ee
where we denoted $F(k)=F_+(k) + F_-(k)$. It is interesting to notice that
having set a restriction on the topology of the spacetime we cannot 
distinguish between particles and anti-particles, keeping in mind Feynman's
Principle of Reinterpretation and interpreting values of f(x,p) in the
area $p_0{}<0$ ("lower" half of mass shell) as the distribution function
of anti-matter. On one hand, having $F_-$ in energy-momentum tensor may
violate dominant energy condition but, on the other hand, it appears to be
more consistent and complete. Nature is symmetric and beautiful and so must
be the equations that describe it however naive this may sound. One can 
derive second order equation for R(t) from (1.7) and (1.8):
\be
\ddot R=-{2\over R} - 2{\dot R^2\over R} +
{1\over 3R^4} \int {F(k) k^2 d^3k \over \sqrt[]{m^2 + k^2/R^2}} +
{m^2 \over 2R^2} \int {F(k) d^3k \over \sqrt[]{m^2 + k^2/R^2}}
\ee
It can be easily shown that the constraint equation (1.10) is satisfied.
Also, it is obvious (in this particular case) that if $R(t)$ satisfies 
(1.18) for arbitrary t then it satisfies (1.19) as well which allows one 
to consider (1.18) as a main dynamical equation. The "maximal" set of 
solutions of (1.19) is, however, larger than that of (1.18) but we shall 
not be concerned with losing some of the solutions for now.

\newsection{Analytical Solution}

In this section we consider a special case of a Universe consisting of
massless m=0 particles. Using dimensionless variables (x,$\tau$) we have:
\ba
R=R(0)x,\nonu
t=R(0)\tau,\nonu
\dot x \equiv {dx\over{d\tau}},\nonu
\epsilon{}={1\over{3}}R(0)^2\rho{}(0)
\ea
Equation (1.17) then becomes:
\ba
\dot x^2=-1+{\epsilon{}\over{}x^2},\nonu
x(0)=1
\ea
If $\epsilon{}\geq{}1$ Cauchy problem (2.2) has a real solution:
\be
x(\tau{})=\sqrt[]{\epsilon{}-(\tau{}-\sqrt[]{\epsilon{}-1})^2}
\ee
Going back to (t,R) variables we have:
\be
R(t)=R(0)\sqrt[]{{R(0)^2\rho (0)\over 3}-\
\Bigl(t/R(0)-\sqrt[]{{R(0)^2\rho (0)\over 3}-1}\Bigr)^2}
\ee
Using the above expression one can evaluate the age of the Universe 
($T_{bb}$) and current value of the Hubble constant:
\ba
R(0)\simeq 1.25\cdot 10^{28} cm,\nonu
\rho (0)\simeq 1.1\cdot 10^{-57} cm^{-2},\nonu
\epsilon \simeq 1.459,\nonu
T_{bb} \simeq -7.09 \cdot 10^9 yrs,\nonu
T_{tot} \simeq 31.7 \cdot 10^9 yrs,\nonu
H={\dot R(0) \over R(0)} \simeq 5.304 \cdot 10^{-29} cm^{-1}
\ea
As follows from these calculations even under assumption m=0 the results
are reasonable ie they lie within the "standard" bounds predicted by
hydrodynamical Friedmann-Robertson-Walker cosmological model. More complicated
case of $m\geq{0}$ can serve as a topic for next paper on the subject.

{\bf Acknowledgements.}\nonu
This work was done several years ago (1992) when author was an undergraduate 
student of Yerevan State University (Armenia) and it remained in the form 
of ``half-forgotten manuscript'' until recently I read an article by  
Rendall [1] which woke up my interest to the Einstein-Vlasov system 
and incentive to ``dig out old papers''.

{\bf References.}\nonu
[1] A. D. Rendall.: An introduction to the Einstein-Vlasov system.
gr-qc/9604001.\nonu
[2] Arnowitt R., Deser S., Misner C.W., in Gravitation: Introduction
to Current Research, ed. L. Witten, New York, Wiley, 1962.

\newpage
\newsection{\bf Appendix A: ADM Representation of Einstein-Vlasov system}

Metric tensor:
\be
g_\mu{}_\nu{}=\pmatrix{
-\alpha{^2}+\beta{_i}\beta{^i} & \beta{}_i      \cr
\beta{_i}                      & \gamma{_{i k}} \cr
}
\ee
Energy-momentum tensor:
\be
T_{\mu \nu}(x)= {2 \over \sqrt[]{-g}} \int f(x,p)p_\mu p_\nu \
\delta (p^2+m^2) d^4p
\ee
Measure of integration in the phase space:
\be
d\mu={2\over \sqrt[]{-g}}\delta (p^2+m^2) d^4p
\ee
Dynamical equations ($c=1,G=1/8\pi$):
\be
\partial{_t}\gamma{_{i j}}=-2\alpha{}K_{i j}+D_i\beta{_j}+D_j\beta{_i}
\ee
\ba
\partial{}_tK_{i j}=-D_i{}D_j{}\alpha{} + \alpha{}(R_{i j}-
2K_i{}_l{}K^l{}_j{} + KK_{i j} - S_{i j} + \nonu
{1\over{2}}\gamma{_{i j}}[\gamma{^{k l}}T_{k l}-\rho{_m}])+\
\beta{^l}D_l{}K_{i j} + K_{l i}D_j{}\beta{^l} + K_{l j}D_i{}\beta{^l}
\ea
\be
\partial{_t}F_\pm=\hat L_\pm F_\pm
\ee
Constraint equations:
\be
R+K^2-K_{i k}K^{i k}=2 \rho{_m}
\ee
\be
D_j{}(K_i{}^j{}-\delta{_i{}^j{}}K)=S_i{}
\ee
The expression for Liouville operators can be obtained from the equations for
geodesics:
\ba
\hat L_\pm = {1\over{}g^{00}p_0^{\pm}+g^{0k}p_k}\
\biggl\{\Bigl(g^{i0}p_0^{\pm}+g^{ik}p_k\Bigr){\partial \over \partial{x^i}} - \
{1\over{2}}\Bigl({\partial g^{00} \over \partial{x^i}}p_o^{\pm 2} + \nonu
2 {\partial g^{0k} \over \partial{x^i}}p_0^{\pm}p_k + \
{\partial g^{kl} \over \partial{x^i}}p_k p_l\Bigr) \
{\partial \over \partial{p_i}}\biggr\}
\ea
We use the following standard notations:
\ba
F_\pm = f(x,p) \mid _{p_0=p_0^{\pm}},\nonu
n_\mu{}(-\alpha,0,0,0),\nonu
n^\mu{}(1/\alpha,-\beta{^i}/\alpha{}),\nonu
\rho{_m}=T_{\mu \nu} n^\mu{} n^\nu{},\nonu
S_i{}=-T_{\mu \nu} n^\mu{} P^\nu{}_i{},\nonu
S_i{}_k{}=T_\mu{}_\nu{}P^\mu{}_i{}P^\nu{}_k{},\nonu
P^\mu{}_\nu{}=\delta{^\mu{}_\nu{}}+n^\mu{}n_\nu{},\nonu
K=\gamma{^{i k}}K_{i k},\nonu
R=\gamma{^{i k}}R_{i k},\nonu
R_i{}_k{}=\gamma{^l{}_{i k , l}}-
\gamma{^l{}_i{}_l{}_,{}_k{}}+
\gamma{^l{}_i{}_k{}}\gamma{^m{}_l{}_m{}}-
\gamma{^m{}_i{}_l{}}\gamma{^l{}_k{}_m{}},\nonu
\gamma{^i{}_j{}_k{}}={\gamma{^i{}^m{}}\over{}2}
(\gamma{_j{}_m{}_,{}_k{}}+\gamma{_k{}_m{}_,{}_j{}}-
\gamma{_j{}_k{}_,{}_m{}})
\ea

\end{document}